\documentclass[prb,aps,twocolumn,showpacs,preprintnumbers,amsmath,amssymb,
superscriptaddress,notitlepage]{revtex4-1}

\usepackage{graphicx}
\usepackage{dcolumn}
\usepackage{bm}
\usepackage{epstopdf}

scaled\magstep1

\let\a=\alpha \let\b=\beta    \let\d=\delta \let\e=\varepsilon
  \let\h=\eta     \let\l=\lambda
\let\m=\mu    \let\n=\nu         \let\p=\pi    
\let\s=\sigma    \let\f=\varphi \let\c=\chi
   \let\o=\omega
\let\G=\Gamma \let\D=\Delta  \let\L=\Lambda 
         
\let\O=\Omega 

 \def\VV{{\cal V}}
 \def\WW{{\cal W}}
 \def\BBB{{\cal B}}
\def\RR{{\cal R}}

   \def\pp{{\bf p}}
 \def\xx{{\bf x}} \def\yy{{\bf y}} 

\def\kk{{\bf k}}

\def\\{\hfill\break}
\let\==\equiv
\let\io=\infty
\let\0=\noindent
\def\media#1{{\langle#1\rangle}}

\def\const{{\rm const}}
\def\tende#1{\,\vtop{\ialign{##\crcr\rightarrowfill\crcr\noalign{\kern-1pt
    \nointerlineskip} \hskip3.pt${\scriptstyle #1}$\hskip3.pt\crcr}}\,}
\def\otto{\,{\kern-1.truept\leftarrow\kern-5.truept\to\kern-1.truept}\,}

\def\to{\rightarrow}

\def\qed{\hfill\raise1pt\hbox{\vrule height5pt width5pt depth0pt}}

\def\V#1{{\bf#1}}
\def\be{\begin{equation}}
\def\ee{\end{equation}}
\def\bea{\begin{eqnarray}}
\def\eea{\end{eqnarray}}
\def\nn{\nonumber}
\def\pref#1{(\ref{#1})}

\begin{document}

\title{Exact RG computation of the optical conductivity of graphene}
\author{Alessandro Giuliani}
\affiliation{%
Universit\`a di Roma Tre, L.go S. L. Murialdo 1, 00146 Roma,
Italy}
\author{Vieri Mastropietro}%
\affiliation{%
Universit\`a di Roma Tor Vergata, Viale della Ricerca Scientifica
00133 Roma, Italy}

\begin{abstract} The optical conductivity of a system of electrons
on the honeycomb lattice interacting through an electromagnetic
field is computed by truncated exact Renormalization Group (RG)
methods. We find that the conductivity has the universal value
$\pi/2$ times the conductivity quantum up to negligible
corrections vanishing as a power law in the limit of low
frequencies.
\end{abstract}
 \maketitle

Among the remarkable properties of graphene \cite{N}, the optical
conductivity is of special interest. Recent experiments \cite{N1}
found that the conductivity in monolayer graphene is essentially
constant in a wide range of frequencies between the temperature
and the band-width. The observed value of the conductivity is
equal, within experimental errors, to $\s_0=\pi e^2/(2 h)$, a
universal value that only depends on fundamental constants and not
on the material parameters like the Fermi velocity. This fact can
be nicely explained in terms of the standard graphene's model of
massless Dirac particles in 2+1 dimensions: in this case, by
neglecting interactions and disorder, Ref.\cite{L} predicted a
value of the conductivity $\s_\b(\o)$ at inverse temperature $\b$
and frequency $\o$ satisfying $\lim_{\o\to
0}\lim_{\b\to\io}\s_\b(\o)=\s_0$. The inclusion of lattice effects
does not change the value of this limit \cite{SPG}.

This remarkable agreement with a theoretical value computed by
{\it neglecting} many body interaction is, however, surprising and
needs an explanation \cite{K}. Indeed, the strength of the interactions in graphene
is measured by the ratio $\a=\frac{e^2}{\hbar v_0}\sim 2.2$ ($e$ is
the electric charge and $v_0$ the Fermi velocity), which is $300$
times larger than the usual fine structure constant. The effects
of the interactions are clearly seen in experiments on the Fermi
velocity \cite{E}. Therefore, why is not there an essential many
body renormalization of the optical conductivity, too?

On the theoretical side, a theorem proved in Ref. \cite{GMPcond}
establishes that the conductivity of electrons hopping on the honeycomb
lattice and interacting via a weak Hubbard interaction is equal to
$\s_0$ in the limit $\o\to 0$. Note that, even if dimensionally
irrelevant, the interaction can produce finite many body
renormalizations: for instance, the Fermi velocity is {\it
renormalized} by the interaction. Therefore, the universality of the
conductivity is a non trivial statement, following from an exact cancellation
of  all the many body corrections.

It is, however, believed that the interaction in clean suspended
graphene is not at all short-ranged as the Hubbard interaction (no
screening), so that a more realistic description of the clean
system requires the inclusion of the long-ranged electromagnetic
(e.m.) interactions. In the case of static Coulomb interactions,
Ref.\cite{V3} predicted a logarithmic renormalization of the Fermi
velocity, namely $ v(q)=v_0\big(1+\frac{\a}{4}\log
{\frac{\e}{q}}\big)$ where $q$ is the momentum measured from the
Fermi points and $\e$ is the bandwidth. First attempts to include
the effects of a Coulomb potential on the conductivity \cite{M1}
led to the conclusion that the interaction radically changes its
behavior, that is $\lim_{\o\to 0}\s(\o)=0$, where $\s(\o)$ is the
conductivity in the limit of zero temperature. Later, Ref.
\cite{SS1,H0} obtained the qualitatively different result
$\lim_{\o\to 0}\s(\o)=\s_0$, based on
scaling arguments. In particular, Ref. \cite{H0} found the formula
%
%
\be\s(\o) = \s_0\big[1+O(\frac{1}{\log( \e/
\o)})\big]\;.\label{s2}\ee
Note that the inverse logarithmic correction in Eq.(\ref{s2}) is a consequence
of the logarithmic divergence of the Fermi velocity, and should be read as $O(\a v_0/v(\o))$.
As pointed out in Ref.\cite{K}, this correction is in general
{\it larger} than the experimental error \cite{N1}.
Ref.\cite{M2,SS2} proposed that the way out from this apparent
contradiction should be found
 in the constant in front of the inverse log
corrections, whose correct value should be much smaller than the
one computed in Ref.\cite{H0}. However, Ref.\cite{H1} raised
objections against the new value proposed in Ref.\cite{M2,SS2},
because the regularizations used in these works can produce
unphysical results. The disagreement between the big (inverse log)
corrections to the conductivity and the experimental data
suggested \cite{K} to phenomenologically postulate a Fermi liquid
description of the interacting system: this assumption implies
that the universal conductivity is reached at low frequencies
polynomially fast (i.e., as $\sim \o^2$) but is in contrast with
the experiments in Ref.\cite{E}.

Eq.\pref{s2} was derived by assuming that the electrons interact
via a static Coulomb interaction: however, the logarithmic
increase of the Fermi velocity shows that the assumption of
instantaneous interactions becomes unphysical at very low energy
scales \cite{Voz}. Therefore, the use of Eq.\pref{s2} and of the divergence of the Fermi velocity
to predict the universality of the conductivity as $\o\to 0$ is
questionable. The unbounded increase of the Fermi velocity is absent in the case
that the interaction with the e.m. field is introduced via
the Peierls substitution in order to preserve gauge invariance. It
is well known\cite{GGV,GMPgauge} that in this case the Fermi
velocity stops flowing at the speed of light $c$ and Lorentz
symmetry spontaneously emerges in the infrared. We compute the
optical conductivity at imaginary frequency $\o$ in a lattice gauge invariant model for graphene
using {\it
truncated exact RG methods}. We find an expression that is
qualitatively different from Eq.\pref{s2}, namely \cite{footnote}
\be \s(\o)= \s_0\big[1+O(\frac{\o}{\e}\log\frac{\e}{\o})\big]\;,\label{1}
\ee
which is very close to the universal one at low frequencies, up to a really negligible
power law correction, compatible with the experimental results in
Ref. \cite{N1}. The $\o\log\o$ dependence of the correction is not necessarily optimal,
it may just be a byproduct of our estimates.

We derived Eq.\pref{1} under the assumption that the
values of the bare parameters are sufficiently close to the
infrared fixed point (i.e., the bare Fermi velocity $v_0$ is
sufficiently large). The extension of its validity to real frequencies
and to a larger range of parameters, including those measured
in actual graphene's samples, requires a microscopic justification
that is quite difficult in view of the strength of interactions in graphene \cite{Rscience};
of course, this is a caveat that applies to all the
approaches based on expansions, resummations and truncations.
In any case, it is reassuring to see that encoding a fundamental
physical principle like gauge invariance into the model is sufficient
to obtain results that are in good
qualitative agreement with the experimental data, in particular with
the observed dramatic increase of the Fermi velocity and with the
universality of the conductivity up to negligible power law
corrections at low frequencies.

The model we consider was defined in detail in Ref.\cite{GMPgauge}.
Let us just remind here the main definitions. The grand-canonical
Hamiltonian at half-filling is $H=H_0+H_C+H_A$, with
$$ H_0=-t\sum_{\substack{\vec x\in\L_A \\
j=1,2,3}}\sum_{\s=\uparrow\downarrow} a^{+}_{\vec x,\s}
b^{-}_{\vec x + \vec \d_j,\s} e^{ie\int_0^1\vec\d_j\cdot\vec
A(\vec x+s\vec\d_j,0)\,ds}
 + c. c. $$
the gauge invariant nearest neighbor hopping term (here $t$ is the hopping strength, $\vec\d_j$ the
nearest neighbor vectors and $a^\pm, b^\pm$ the creation/annihilation operators of electrons sitting
at the sites of the $A$ or $B$ sublattice of the honeycomb lattice),
$$H_C=\frac{e^2}2\sum_{\vec x,\vec y\in\L_A\cup\L_B}
(n_{\vec x}-1)\f({\vec x}-{\vec y})(n_{\vec y}-1)\;,$$
where $e$ is the electric charge, $\hat\f_{\vec p}$ is a
regularized version of the static Coulomb potential and $n_{\vec
x}$ the electron number at site $\vec x$. Finally, $H_A$ is the
energy (in the presence of an ultraviolet cutoff) of the
three-dimensional photon field $\underline A=(\vec A,A^3)$ in the
Coulomb gauge. Units are fixed in such a way that the speed of
light $c=1$. Note that the interaction with the quantum e.m. field
is introduced via the Peierls substitution in order to preserve
Gauge invariance.

Proceeding as in Ref.\cite{GMPcond}, where we computed the conductivity in the case of short range interactions, we define a
``space-time" three-components vector $\hat J_{\vec p,\m}$,
$\m=0,1,2$, with
\be \hat J_{\vec p,0}=e\sum_{\substack{\vec x\in\L_A\\
\s=\uparrow\downarrow}} e^{-i\vec p\vec x}a^+_{\vec x,\s}a^-_{\vec
x,\s}+
\sum_{\substack{\vec x\in\L_B\\
\s=\uparrow\downarrow}}e^{-i\vec p\vec x}b^+_{\vec x,\s} b^-_{\vec
x,\s}\;, \label{3.rho}\ee
the density operator and $\hat J_{\vec p,1}, \hat J_{\vec p,2}$ the two components of the {\it paramagnetic current}
\be\vec J_{\vec p}=iet\sum_{\substack{\vec x\in\L\\
\s,j}}\,e^{-i\vec p\vec x} \vec \d_j\h^j_{\vec p}\big(a^+_{\vec
x,\s}b^-_{\vec x+\vec \d_j,\s}- b^+_{\vec x+\vec \d_j,\s}
a^-_{\vec x,\s}\big)\nn \ee
where $\h^j_{\vec p}=(1-e^{-i\vec p\vec \d_j})/(i\vec p\vec
\d_j)$. Let also $\pp=(\o,\vec p)$, with $\o$ the
Matsubara frequency, and $\hat K_{\m\n}(\pp)$ is the current-current response function,
i.e., the Fourier transform of
$\lim_{\b\to\infty}\media{J_{\xx,\m};J_{\yy,\n}}_{\b}$.

We are interested in the
{\it conductivity}, defined via Kubo formula as \cite{SPG,GMPcond} (here
$l,m=1,2$):
\be \s_{lm}(\o )= -\frac{2}{3\sqrt3}\frac1{\o } \Big[\hat
K_{lm}(\o,\vec 0)+\hat \D_{lm}(\vec 0)\Big]\;,\nn\ee
where $3\sqrt3/2$ is the area of the
hexagonal cell of the honeycomb lattice and
$$\hat \D_{lm}(\vec p)=\lim_{\b,L\to\infty}\frac1{L^2}
\sum_{\substack{\vec x\in\L\\ j=1,2,3}} (\vec \d_j)_l(\vec \d_j)_m|\h^j_{\vec p}|^2
\media{\D_{\vec x,j}}_\b\;,$$
with $\D_{\vec x,j}=-e^2 t\sum_{\s}(a^+_{\vec x,\s}b^-_{\vec
x+\vec\d_j,\s}+ b^+_{\vec x+\vec\d_j,\s}a^-_{\vec x,\s})$ the
{\it diamagnetic tensor}.

The current-current response function can be computed via the generating functional that,
in the Feynman gauge, reads
\be e^{\WW_{h^*}(J,\l)}=\int P(d\psi)\int P_{h^*}(dA)e^{\VV(A+J,\psi)+(\psi,\l)}\label{por}
\ee
which has been studied in great detail in Ref.\cite{GMPgauge}. In Eq.(\ref{por}): (i)
$\psi^\pm_{\kk,\s}$ are Grassman spinors (of the form $\psi=(a,b)$, with $a$ and $b$ the electron
fields associated to the two sublattices of the honeycomb net) and $P(d\psi)$ is the fermionic gaussian integration with propagator
\be g(\kk)= -\frac1{Z_0}\left(\begin{array}{cc} ik_0 & v_0 \O^*(\vec k)
\\ v_0 \O(\vec k) & ik_0 \end{array}\right)^{\!\!-1}\;\label{vo}\ee
where $Z_0=1$ is the free wave function renormalization,
$v_0=\frac32t$ is the free Fermi velocity and $\O(\vec k) =
\frac23\sum_{j=1,2,3}e^{i\vec k(\vec \d_j - \vec\d_1)}$ is the
complex dispersion relation. Note that $g(\kk)$ is singular only
at the Fermi points
$\pp_F^{r}=(0,\frac{2\p}{3},r\frac{2\p}{3\sqrt3})$, where $r=\pm$
is the {\it valley index}. Moreover, $A_\m(\pp)$, $\m=0,1,2$,  are
the Fourier transform of real gaussian variables and $P_{h^*}(dA)$
is the gaussian integration with propagator $w_{\m\n}(\pp)=
\d_{\m\n}(2|\pp|)^{-1}\c_{[h^*,0]}(|\vec p|)$, where
$\chi_{[h^*,0]}$ is a smooth compact support function that acts
both as an ultraviolet cutoff on scale $|\pp|\sim 1$ {\it and} as
an infrared cutoff on scale $|\pp|\sim 2^{h^*}$ (to be eventually
removed). Finally $\VV$ is the interaction whose explicit form can
be easily inherited from $H$ \cite{GMPgauge}. The current-current
response function can be obtained by taking the limit
$h^*\to-\infty$ and by deriving twice with respect to the external
field $J$ and then setting $J=\l=0$. Field-field correlations or
field-current correlations can be obtained similarly, by suitably
deriving with respect to the external fields $\l$ and/or $J$. Note
that in writing the generating functional as in Eq.(\ref{por}) we
exploited gauge invariance and, more precisely, the equivalence
between the Feynman and the Coulomb gauges. Another crucial
consequence of gauge invariance is the following equation
\be 0 = \frac{\partial}{\partial\hat \a_{\pp}} \WW(\Phi,J +
\partial\a,\l e^{ i e \a}) \Big|_{\hat \a=0}\;.\label{WI}\ee
By performing derivatives with respect to the external
fields, this equation also implies a sequence of exact lattice {\it Ward Identities}, valid for each finite
choice of the cutoff scale $h^*$. In
particular, proceeding as in Ref.\cite{GMPcond} and defining $p^0=-i\o$, the current-current response function satisfies the Ward Identities $\sum_{\m=0}^2p^\m\hat K_{\m 0}(\pp)=0$ and, for $j=1,2$,
\be \sum_{\m=0}^2p^\m\hat K_{\m
j}(\pp)=-\sum_{i=1}^2 p_i\hat\D_{ij}(\vec p)\;.\label{wi2}\ee
An immediate consequence of Eq.(\ref{wi2}) {\it and} of the continuity of $\hat K_{\m,\n}(\pp)$
in $\pp=\V0$ (proved at all orders of renormalized perturbation theory \cite{GMPgauge}) is that,
for $i,j\in\{1,2\}$,
\be \s_{ij}(\o)= -\frac{2}{3\sqrt3}\frac1{\o} [\hat
K_{ij}(\o,0)-\hat K_{ij}(0,\vec 0)]\;,\label{conc}\ee
see \cite{GMPcond} for the simple argument leading to Eq.(\ref{conc}).

The generating function \pref{por} can be computed by exact
RG methods \cite{GMPgauge}, which allowed us to prove that
the response functions can be written in terms of a renormalized perturbation theory that
is finite at all orders in the effective coupling constants, with explicit bounds on the $n$-th order
contributions. In particular, after the integration of the degrees of freedom corresponding to
momenta larger than $2^h$, $h<0$, we rewrite (setting for simplicity $\l=0$): $ e^{\WW_{h^*}(J,0)} =$
\be=\int \prod_{r=\pm}P(d\psi_r^{(\leq h)})P(dA^{(\leq
h)})e^{\VV^{(h)}(\sqrt{Z_h}\psi^{(\leq h)},\, A^{(\leq
h)}+J)}\;,\nn\ee
where $P(d\psi_{r}^{(\le h)})$ and $P(d A^{(\le h)})$ have
propagators
\be\hat g^{(\leq h)}_{r}(\kk') = - \frac{\chi_h(\vec k')}{Z_h}\begin{pmatrix} ik_0 &
v_h\O^*(\vec k'+ \vec p_F^{\,r})\\  v_h\O(\vec k'+ \vec
p_F^{\,r}) & ik_0
\end{pmatrix}^{\!\!\!-1}\label{proph}\ee
and $ w_{\m\n}^{(\leq h)}(\pp) =\d_{\m\n}(2|\pp|)^{-1}\c_{[h^*,h]}(\vec
p)$, where: (i) $\chi_h(\vec k')$ is a smooth
cutoff function vanishing for momenta larger than $|\vec k'|\sim 2^h$; (ii)
$\c_{[h^*,h]}(\vec p)=\c_h(\vec p)-\c_{h^*}(\vec p)$;
(iii) $Z_h, v_h$ are the
effective wave function renormalization and Fermi velocity at
scale $h$. Moreover $\VV^{(h)}$ is the {\it effective potential},
expressed by a sum of monomials in $\psi^{(\le h)},A^{(\le h)}$ of any
degree:
\bea &&\VV^{(h)}(\sqrt{Z_{h}}\psi^{(\leq h)}, A^{(\leq h)}) =\int\frac{d\pp}{(2\p)^3}
\big[Z^{(\m)}_{h}e \hat
\jmath^{(\leq h)}_{\m,\pp}\hat A^{(\leq h)}_{\m,\pp}
 - \nn\\
 &&2^{h}\n_{\m,h}\hat A^{(\leq h)}_{\m,-\pp}\hat A^{(\leq
h)}_{\m,\pp}\big]+\RR\VV^{(h)}(\sqrt{Z_{h}}\psi^{(\leq h)},
A^{(\leq h)})\label{3.14b} \eea
where $\RR\VV^{(h)}(\sqrt{Z_{h}}\psi^{(\leq h)}, A^{(\leq h)})$ is
the {\it irrelevant part} of the effective potential (sum of all
the terms with more than three fields) and
\be \hat\jmath^{(\leq h)}_{\m,\pp} := \frac{i}{\b
L^2}\sum_{r,\s,\kk'}\, \hat\psi^{(\leq
h)+}_{\kk'+\pp,\s,r}\G^{\m}_{r}(\vec k'+\vec p_F^{\,r})\hat\psi^{(\leq
h)-}_{\kk',\s,r}\;,\label{3.15aa} \ee
with $\G^0(\vec k)=\openone$ and
$$\G^i(\vec k)={2\over 3}\sum_{j=1}^3 (\vec\d_j)_i\hskip-.1truecm
\begin{pmatrix} 0 &\hskip-.1truecm i e^{-i\vec k(\vec \d_j-\vec\d_1)}\\
-i e^{i\vec k(\vec\d_j-\vec\d_1)}&0\end{pmatrix}\;.$$
We can summarize the previous discussion by saying that after the integration of the degrees of freedom
corresponding to momenta $\ge 2^h$, we get an effective
theory that is qualitatively very similar to the original one, modulo the renormalization of a finite
number of effective parameters, namely the Fermi velocity $v_h$, the wave function renormalization
$Z_h$, the vertex function $Z_h^{(\m)}$ and the photon mass $\n_{\m,h}$. The discrete rotational symmetries of the model imply that $Z_h^{(1)}=Z^{(2)}_h$ and $\n_{1,h}=\n_{2,h}$.

These parameters verify suitable flow equations well defined at
all orders in the renormalized expansion: this is an instance of
the renormalizability at all orders of the model. Note that the
effective charges at scale $h$ are: $e_{0,h}=eZ_h^{(0)}/Z_h$ and
$e_{i,h}=e Z_h^{(i)}/(Z_h v_h)$, for $i\in\{1,2\}$. Thanks to the
WIs induced by Eq.(\ref{WI}), we proved in Ref.\cite{GMPgauge}
that $\n_{\m,h}=O(e^2 2^h)$ (i.e., the effective photon mass
vanishes in the infrared) and that the beta function for the
effective charge is asymptotically vanishing, i.e.,
\bea && e_{0,h}=e\frac{Z^{(0)}_h}{Z_h}=e\big(1+O(e^2_{\m,h})\big)\;,\nn\\
&& e_{i,h}=e\frac{Z^{(i)}_h}{Z_h
v_h}=e\big(1+O(e^2_{\m,h})\big)\label{1a}\eea
and $e_{0,-\io}=e_{1,-\io}=e_{2,-\io}$. Moreover, see
\cite{GMPgauge}, the wave function renormalization diverges in the infrared,
while the effective Fermi velocity increases up to the speed of light, both approaching their limits
with an anomalous power law:
\be Z_h\sim 2^{-\h h}\quad\quad 1-v_h\sim 2^{\tilde\h h}\ee
with $\h=\frac{e_{-\io}^2}{12\pi^2}+O(e_{-\io}^4)$ and $\tilde\h=
\frac{2e_{-\io}^2}{5\pi^2}+O(e_{-\io}^4)$ the two critical exponents.

The above integration procedure leads to an expansion of the
conductivity in terms of powers of $e_{\m,h}$; such renormalized
expansion is a resummation of the naive perturbative expansion in
$e$. It must be stressed that there is a big difference between these two expansions:
while the one in $e_{\m,h}$ is order by order finite (with explicit bounds on the growth
of the $n$-th order contributions \cite{GMPgauge}), the naive one in $e$ is plagued by
$O(\log^n\o)$ divergences at order $n$. Therefore, the truncation of the renormalized expansion is
expected to give much more accurate predictions than the naive one.

\begin{figure}[hbtp]
\centering
\includegraphics[width=.4\textwidth]{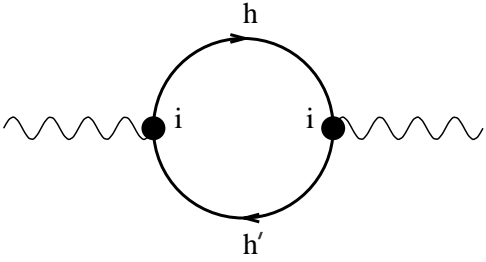}
\caption{The one-loop bubble diagram contributing to the longitudinal conductivity $\s_{ii}(\o)$.
The labels
$h,h'$ indicate the scale of the two loop propagators, which depend on the dressed Fermi velocities
$v_h,v_{h'}$ and on the effective wave function renormalizations $Z_h, Z_{h'}$.
A summation over $h,h'$ is understood.
The big dots correspond to dressed vertex functions
$Z_{\bar h}^{(i)}$, with $\bar h=\max\{h,h'\}$. The external momentum flowing in the wavy lines is
$\pp=(\o,\vec 0)$.}\label{figkekbis}
\end{figure}

By truncating the exact RG expression for the conductivity at one
loop, we get contributions from the bubble diagrams in Fig.\ref{figkekbis},

which give (defining $\s_0=\frac{\p}{2}\frac{e^2}{h}$):
\bea &&
\frac{\s_{ii}(\o)}{\s_0}=\frac{16}{3\sqrt3}\frac1{\o}\sum_{h,h'\le 0}^{r=\pm}\int\frac{d
k_0}{2\p} \int_{\BBB}\frac{d\vec k'}{|\BBB|}\frac{(Z^{(i)}_{\bar h})^2}{Z_h Z_{h'}} \cdot\label{fon}\\
&& \cdot
{\rm Tr}\Big\{\G^i_r(\vec k') g^{(h)}_r(\kk')\G^i_r(\vec k')\big[ g^{(h)}_r(\kk'+(\o,\vec 0))-
g^{(h)}_r(\kk')\big]\Big\}\;.\nn\eea
where: (i) $\BBB$ is the first Brillouin zone and $|\BBB|=8\p^2/(3\sqrt3)$ its area; (ii)
$\bar h=\max\{h,h'\}$; (iii) $\G^i_r(\vec k')=\G^i(\vec k'+\vec p_F^{\, r})$; (iv) $g^{(h)}_r$ is the effective
propagator on scale $h$, given by the same expression as Eq.(\ref{proph}) with $\c_h(\vec k')$ replaced
by the smooth compact support function $f_h(\vec k'):=\c_{h}(\vec k')-\c_{h-1}(\vec k')$,
which is non vanishing only if $\vec k'$ is on scale $h$, i. e., $2^{h-1}\le |\vec k'|\le
2^{h+1}$. Note that the effective parameters $Z^{(i)}_h, Z^h, v_h$ entering Eq.(\ref{fon})
are all functions of $e$: if we expanded them in $e$ we would recover infinitely many graphs of
the naive
perturbation theory, all plagued by logarithmic divergences. Note also
that Eq.\pref{fon} is not simply the ``bubble graph" with the dressed
propagator and vertices: e.g., if one thinks of the dressed propagator with momentum $\kk$
as being obtained by
resummations of the chain of self-energies, one has to take into account that the scales of the
momenta flowing inside such self-energy sub-diagrams are higher than the scale of $\kk$,
according to the rules of exact RG (which avoid the problem of overlapping divergences and, correspondingly, the emergence of
$n!$ factors at higher orders).

The computation of Eq.(\ref{fon}) can be explicitly performed, by
making use of Eq.(\ref{1a}) and by carefully exploiting symmetry
cancellations that make the apparent logarithmic divergence of
Eq.(\ref{fon}) finite, see Appendix \ref{app1} for details.  The
result is Eq.(\ref{1}).

Our analysis is based on a truncation of the exact RG equations, and
the question of how to generalize it to the full RG expansion is a very
interesting and important theoretical problem; so far, we succeeded in
performing the full RG computation only in the case of short range
interactions \cite{GMPcond}. Another important open problem is
to understand the analytic extension of the conductivity to real frequencies.

In conclusion, we considered a model of electrons on the honeycomb
lattice interacting via a quantized photon field, previously
investigated in Ref.\cite{GMPgauge}. The coupling with the e.m.
field is introduced via the Peierls substitution in order to preserve gauge
invariance. We showed that at low frequencies the
conductivity is equal to the universal value $\s_0$ up to
corrections $O(\o\log\o)$, which are much smaller than the
$1/\log\o$ corrections found for static Coulomb interactions. Our
results are a priori valid close to the infrared fixed point and
the extension of their validity to a larger range of bare
parameters (including those measured in actual graphene's samples)
is based on a phenomenological assumption. Still, it is reassuring to see that
it is enough to encode gauge invariance in a microscopic model for clean graphene
to recover good qualitative agreement of the predictions with the experimental data.
\\

{\bf Acknowledgements} We acknowledge financial support from the ERC Starting Grant CoMBoS-239694.

\appendix
\section{Computation of the conductivity}\label{app1}

Using the definition of $g^{(h)}_r$ into Eq.(\ref{fon}), we can rewrite
(shifting the momenta by $\vec p_F^{\, r}$ and neglecting for consistency terms of order $O(e^4)$, which
should be combined with the two-loops contributions)
\bea && \frac{\s_{ii}(\o)}{\s_0}=\frac4{\p^2}\frac1{\o}
\sum_{h,h'\le 0}^{r=\pm}\int\frac{d k_0}{2\p} \int_{\BBB}d\vec
k'\frac{(Z_{\bar h}^{(i)})^2
}{Z_h Z_{h'}}\frac{f_h(\vec k')f_{h'}(\vec k')}{k_0^2+v_h^2|\O_r(\vec k')|^2}\cdot\nn\\
&&\cdot\Biggl[\frac{-k_0(k_0+\o)|a_{i,r}(\vec k')|^2+v_hv_{h'}{\rm
Re}\big(\O_r^2(\vec k')a^2_{i,r}(\vec
k')\big)}{(k_0+\o)^2+v_{h'}^2|\O_r(\vec k')|^2}+\nn\\
&&+ \frac{k_0^2|a_{i,r}(\vec k')|^2-v_hv_{h'}{\rm Re}\big(\O_r^2(\vec
k')a^2_{i,r}(\vec k')\big)}{k_0^2+v_{h'}^2|\O_r(\vec k')|^2}
\Biggr]\;,\label{cond_ii}\eea
where $\O_r(\vec k')=\O(\vec k'+\vec p_F^{\,r})$ and $$a_{i,r}(\vec k')=
\frac23\sum_{j=1}^3 (\vec\d_j)_i i e^{-i(\vec k'+\vec p_F^{\,r})(\vec \d_j-\vec\d_1)}\;.$$ Note that
$\O_r(\vec k')=ik_1'+r
k_2'+O(|\vec k'|^2)$, $a_{1,r}(\vec k')=i+O(|\vec k'|)$ and $a_{2,r}(\vec k')=-r+O(|\vec k'|)$.
The ``relativistic approximation" consists in replacing $\O_r(\vec k')$, $a_{1,r}(\vec k')$
and $a_{2,r}(\vec k')$ in Eq.(\ref{cond_ii}) by $ik_1'+r
k_2'$, by $i$ and by $-r$, respectively. By performing this replacement, it becomes apparent that
the r.h.s. of Eq.(\ref{cond_ii}) behaves dimensionally as $\frac1\o\int d^3k'[\frac1{(k'+\o)^2}-\frac1{(k')^2}]$,
which is logarithmically divergent as $\o\to 0$. In order to prove the finiteness of $\s_{ii}(\o)$ in the low
frequency limit, it is necessary to exploit cancellations, which follow from the use of a Ward Identity
combined with an
essentially explicit computation (using the residues' theorem to integrate $k_0$ out)
of the r.h.s. of Eq.(\ref{cond_ii}).

Let $\e$ be a small but finite fraction of the bandwidth (say,
$\e=t/10$) and let us distinguish the contributions to the
integral coming from the region $v_0|\O_r(\vec k')|\ge \e$ from those
$v_0|\O_r(\vec k')|\le \e$. The former correspond to non-singular
contributions, which can be estimated as follows: we expand in
Taylor series the expression in square brackets up to $O(\o^2)$;
we note that the term linear in $\o$ is vanishing by parity in
$k_0$; we bound dimensionally the term quadratic in $\o$ as:
\be (\const.)\frac1{\o}\int dk_0\int_{\e/v_0}^1 \hskip-.2truecm dk'\, k'
\frac{v_0^2\o^2}{(k_0^2+v_0^2(k')^2)^2}\le (\const.)\frac{\o}{\e}\;.\ee
Let us now look at the the terms coming from the region $v_0|\O_r(\vec
k')|\le \e$. Note
that the contributions from $r=+$ or $r=-$ are equal among each other, thanks to the symmetry under
valley exchange. We introduce some shorthands: we define $\D_h=v_h|\O_+(\vec k')|$ and $W_{h,h'}=
v_hv_{h'}{\rm Re}\big(\O_+^2(\vec k')a^2_{i,+}(\vec k')\big)$, so
that, by performing the integral over $k_0$ using the residues' theorem,
we can write the contribution to the conductivity coming from the region $v_0|\O_r(\vec
k')|\le \e$ as
\bea &&\frac8{\p^2}\frac1{\o} \int_{|\vec k'|\le \e/v_0}\hskip-.2truecm d\vec
k'\sum_{h,h'\le 0} \frac{(Z_{\bar h}^{(i)})^2}{Z_h
Z_{h'}}f_h(\vec k')f_{h'}(\vec k')\cdot\nn\\
&&\cdot\Big\{\frac{\D_h(\D_h-i\o)|a_{i,+}(\vec k')|^2+W_{h,h'}}{2\D_h\big[\o^2+2i\o\D_h-(\D_h^2-\D_{h'}^2)\big]}+\label{fon2}\\
&&+\frac{\D_{h'}(\D_{h'}+i\o)|a_{i,+}(\vec k')|^2+W_{h,h'}}{2\D_{h'}\big[\o^2-2i\o\D_{h'}+(\D_h^2-\D_{h'}^2)\big]}+\nn\\
&&+ \frac{\D_h^2|a_{i,+}(\vec k')|^2+W_{h,h'}}{2\D_h(\D_h^2-\D_{h'}^2)}-
\frac{\D_{h'}^2|a_{i,+}(\vec k')|^2+W_{h,h'}}{2\D_{h'}(\D_h^2-\D_{h'}^2)}\Big\}
\;.\nn\eea
Let us first consider in Eq.\pref{fon2} the terms with the integrand proportional to
$W_{h,h'}$, which can be rewritten as (defining $V_{h,h'}=\D_{h}+\D_{h'}$)
\be \int\limits_{|\vec k'|\le \e/v_0}\hskip-.2truecm\frac{d\vec
k'}{\p^2}\hskip-.2truecm\sum_{h,h'\le 0} \frac{(Z_{\bar
h}^{(i)})^2}{Z_h Z_{h'}}\frac{W_{h,h'}}{\D_h\D_{h'}}\frac{-4\o
f_h(\vec k')f_{h'}(\vec
k')}{V_{h,h'}(\o^2+V_{h,h'}^2)}\;.\label{7}\ee
Now note that $W_{h,h'}=v_hv_{h'}\big[(k_2')^2-(k_1')^2+O(|\vec k'|^3)\big]$: therefore, the
term proportional to $(k_2')^2-(k_1')^2$ is zero by symmetry, and we
are left with a contribution dimensionally bounded as (using
that $\D_h$, $\D_{h'}$ and $V_{h,h'}$ behave dimensionally as $\sim k'$ close to the singularity)
\be (\const.)\,\o\int_0^{\e/v_0} \hskip-.3truecm dk'\, k' \frac{v_0(k')^3}{(k')^3(\o^2+v_0^2(k')^2)}\le
(\const.)\,\frac{\o}{v_0}\log\frac{\e}{\o}\;,\label{fo}\ee
for $\o\ll\e$. We are left with the terms obtained by replacing $W_{h,h'}$ with $0$ in Eq.\pref{fon2},
which are given by
\bea &&\frac4{\p^2}\o\int_{|\vec k'|\le \e/v_0}\hskip-.3truecm d\vec k'\sum_{h,h'\le 0}
\frac{(Z_{\bar h}^{(i)})^2}{Z_h Z_{h'}}f_h(\vec k')f_{h'}(\vec
k')\cdot\nn\\
&&\cdot\frac{|a_{i,+}(\vec k')|^2}{ |\O_+(\vec
k')|(v_h+v_{h'})\big[\o^2+(v_h+v_{h'})^2|\O_+(\vec
k')|^2\big]}\;.\nn\eea
Using the Ward Identity Eq.(\ref{1a}) and the rewritings
$\O_+(\vec k')=ik_1'+ k_2'+O(|\vec k'|^2)$, $a_{1,+}(\vec
k')=i+O(|\vec k'|)$ and $a_{2,+}(\vec k')=-1+O(|\vec k'|)$ (valid
close to the singularity $\vec k'=\vec 0$), this is equal (up to
terms coming from the ``non-relativistic" parts of $\O_+(\vec k')$
and $a_{i,+}(\vec k')$, which are bounded as in Eq.\pref{fo}) to
\be\frac4{\p}\o\sum_{h,h'\le 0}\int_0^{\e v_{\bar h}/v_0}\hskip-.3truecm  dk'
\frac{Z_{\bar h}^2}{Z_hZ_{h'}}
\frac{f_h(k')f_{h'}(k')}{(\frac{v_h+v_{h'}}{2v_{\bar
h}})\big[\o^2+(\frac{v_h+v_{h'}}{2v_{\bar h}})^2
4(k')^2\big]}\;.\label{10}\ee
By the compact support properties of $f_h$, the integrand is non vanishing only of $|h-h'|\le 1$,
in which case
\be
\frac{v_h+v_{h'}}{v_{\bar h}}=1+O(e^2)\;,\qquad\frac{Z_{\bar h}^2}{Z_hZ_{h'}}=1+O(e^2)\;,\nn \ee
uniformly in $h$. Therefore, Eq.(\ref{10}) is equal to
\be \int_0^\e \frac{dk'}{\p}\,\frac{4\o}{\big[\o^2+
4(k')^2\big]}=\frac{2}{\p}\arctan(\frac{2\e}{\o})=1-\frac{\o}{\e\p}+O(\o^2)\;,\label{11}\ee
up to terms bounded by $(1-v_0)\o/\e$ and terms bounded {\it
uniformly} in $\o$ by
\be (\const.)e^2\int_0^\e dk'\frac{\o}{(k')^2+\o^2}\le (\const.)e^2\;.\ee
Of course, these correction terms should be neglected, for
consistency, because Eq.\pref{fon} is obtained by truncating the
exact RG expansion at one loop. Putting all together we get
Eq.\pref{1}.


\begin{thebibliography}{999999}
\bibitem{footnote} In Eq.(\ref{1}), by writing
$O(\frac{\o}{\e}\log\frac{\e}{\o})$, we mean that the correction is bounded from above by
 $(\const.)\frac{\o}{\e}\log\frac{\e}{\o}$ for $\o\ll\e$.
\end{thebibliography}
\end{document}